\documentclass[longbibliography]{revtex4-2}

\usepackage{graphicx}
\usepackage{dcolumn}
\usepackage{bm}
\usepackage{siunitx} 
\usepackage{hyperref}
\usepackage{epstopdf}
\usepackage{rotating}
\usepackage{array}
\usepackage{color}
\usepackage{slashed}
\usepackage{multirow}
\usepackage{amsmath}
\usepackage[makeroom]{cancel}
\usepackage{tabularx}
\usepackage{ulem} 
\usepackage{cancel}

\begin{document}

\title{\Large Explicit CP-violating effects on Spin-0 Dark Matter \\ in the I(2+1)HDM with a $Z_2$ symmetry }

\author{A. Dey}
\email[]{atri.dey@physics.uu.se}
\affiliation{Department of Physics and Astronomy, Uppsala University, Box 516, 751 20 Uppsala, Sweden}

\author{J. Hern\'andez-S\'anchez}
\email[]{jaime.hernandez@correo.buap.mx}
\affiliation{Facultad de Ciencias de la Electr\'onica, Benem\'erita Universidad Aut\'onoma de Puebla, Apartado Postal J-48, 72570 Puebla, M\'exico and}
\affiliation{Dual CP  Institute of High Energy Physics, Puebla,  M\'exico }
\affiliation{Facultad de Ciencias, Universidad de Colima, C.P. 28045, Colima, M\'exico}

\author{C. G. Honorato}
\email[]{carlosg.honorato@correo.buap.mx}
\affiliation{Facultad de Ciencias de la Electr\'onica, Benem\'erita Universidad Aut\'onoma de Puebla, Apartado Postal J-48, 72570 Puebla, M\'exico}

\author{S.~Moretti}
\email[]{stefano.moretti@cern.ch}
\affiliation{School of Physics and Astronomy, University of Southampton, Southampton, SO17 1BJ, United Kingdom}
\affiliation{Department of Physics and Astronomy, Uppsala University, Box 516, SE-751 20 Uppsala, Sweden}

\author{T.~Shindou}
\email[]{tetsuo.shindo@icu.ac.jp}
\affiliation{Department of Natural Science, International Christian University, Osawa 3-10-2, Mitaka, Tokyo, 181-8585, Japan}

\date{\today}

\begin{abstract}
{
We investigate the phenomenological consequences of explicit CP-Violating (CPV) effects in the scalar sector of the Inert $(2+1)$ Higgs Doublet Model (I(2+1)HDM), 
a 3-Higgs-Doublet Model (3HDM) in which two inert doublets are odd under a $Z_2$ symmetry while the third is $Z_2$ even and corresponds to the Standard Model (SM) Higgs doublet.
In the so-called `dark democracy' scenario the scalar potential retains two physical phases, 
which turn the four neutral inert states into mixtures of CP eigenstates and, 
in particular, provide different gauge couplings with respect to those in the CP-Conserving (CPC) limit.
Taking the lightest inert state as the Dark Matter (DM) candidate, we select Benchmark Points (BPs) for which the lightest inert scalar provides a subdominant but cosmologically viable DM component, compatible with present direct- and indirect-detection limits. Furthermore, we 
 discuss whether the CPV and the CPC regimes can be told apart at a future $e^+e^-$ collider through the process $e^+e^-\to l^+l^-+\cancel{E}_T$. 
We find that the pseudorapidity of the final-state leptons and their angular separation $\Delta R(l^+,l^-)$ contribute to separate the two regimes and that triple products built out of the initial- and final-state momenta together with the  missing transverse energy ($\cancel{E}_T$) provide a further critical discriminator. 
}

\end{abstract}

\maketitle

\section{Introduction}

Following the 2012 discovery of a scalar particle weighing roughly $125\ GeV$ by the ATLAS and CMS collaborations at the Large Hadron Collider (LHC)  \cite{ATLAS:2012yve,CMS:2012qbp}, significant attention has been focused on studying  its nature and proprieties. 
While its characteristics currently align with those of the SM Higgs boson, this particle could potentially represent the first piece into a more complex scalar sector. 
Despite the absence of direct evidence for Beyond the SM (BSM) physics to date, the SM is fundamentally recognized as an incomplete framework. A primary driver for exploring BSM theories, mainly with extended scalar sectors, is the SM  inability to account for a viable candidate for cold Cold DM (CDM).

While the fundamental identity of DM remains elusive, the $\Lambda-CDM$ cosmological model \cite{Planck:2015fie} dictates that such a particle must be stable over cosmic timescales, non-baryonic and electrically neutral. In addition, to facilitate galaxy formation, it must also be "cold" (i.e., non-relativistic) and  weakly interactive. Among the various theoretical candidates fulfilling these criteria, Weakly Interacting Massive Particles (WIMPs) \cite{Bergstrom:2000pn,Jungman:1995df, Bertone:2004pz} are the most extensively researched, typically spanning a mass range from several GeV to the TeV scale. A viable WIMP must achieve cosmological stability -—often through the imposition of a discrete symmetry -— and undergo a thermal freeze-out process to produce the measured relic abundance  $\Omega_{\rm DM}h^2=0.1199\pm 0.0027$ \cite{Planck:2015fie}.

It is well known that the scalar sector of the SM lacks a suitable WIMP candidate. However, it is possible to extend its scalar sector with the goal of obtaining one (or more) acceptable DM candidates. Specifically, the scalar potential could be enlarged by introducing an additional doublet that preserves a discrete $Z_2$ symmetry and does not acquire a Vacuum Expectation Value (VEV) \cite{Deshpande:1977rw}. This framework, known as the Inert Doublet Model (IDM), extensively documented in recent literature \cite{Ma:2006km, Barbieri:2006dq, LopezHonorez:2006gr}, consists of one active Higgs doublet and one inert scalar doublet, denoted here as I(1+1)HDM, with a vacuum configuration given by $(0,v)$. Consequently, the second doublet (with zero VEV) does not couple to fermions. By construction, this inert doublet is odd under a $Z_2$ symmetry, thereby providing a stable DM candidate. In this way, the I(1+1)HDM serves as a classic example of a Higgs-portal model, where DM annihilation, DM-nucleon scattering cross section, and invisible Higgs decays are governed by the DM-Higgs coupling ($g_{h DM DM}$). However, satisfying all of the aforementioned  experimental constraints simultaneously is challenging. So that, expanding the inert (or dark) dark sector introduces new (co)annihilation channels, which can open up a wider allowed parameter space.

A simple extension of the I(1+1)HDM consists of adding an extra inert $SU(2)$ doublet: this configuration defines the I(2+1)HDM, i.e., a 3HDM with two inert doublets and one active doublet. Following the I(1+1)HDM setup, we consider here the Vacuum Expectation Value (VEV) configuration $(0,0,v/\sqrt{2})$, 
which inherently forbids tree-level Flavor-Changing Neutral Currents (FCNCs). Furthermore, herein, in the presence of one or more discrete symmetries separating the SM from the additional Higgs states,
the lightest neutral members of the inert Higgs sector can be  DM candidates.
Such a model has been proposed and extensively studied in the literature \cite{Keus:2013hya,Keus:2014jha,Keus:2014isa,Keus:2015xya,Keus:2015hva,Cordero-Cid:2016krd,Cordero:2017owj,Cordero-Cid:2018man,Rojas-Ciofalo:2019kar,Aranda:2019vda,Cordero-Cid:2020yba, Hernandez-Sanchez:2020aop, Hernandez-Otero:2022dxd,Hernandez-Sanchez:2022dnn,Dey:2023exa,Dey:2024epo,Belyaev:2026umj,Arroyo-Urena:2026vgf}, in both its CPC and CPV configuration,  in the presence of a $Z_2$, $Z_2\times Z_2$ and $Z_3$ symmetry, considering both DM and collider phenomenological applications.

Herein, we confine ourselves to the case of an I(2+1)HDM with a single $Z_2$ symmetry in presence of (explicit) CP-violation, thereby producing a single DM candidate, of which we will study the properties in collider settings, specifically, at an $e^+e^-$ collider. In particular, we wish to isolate scattering therein processes and (differential) observables that carry an explicit hallmark of CP-violation, i.e., when the latter are notably different from the CPC case. This approach surpasses previous literature (including of ours), which limited itself to look at differences between the two CP setups in the case of integrated observables (like cross sections and asymmetries). 

The plan of the paper is as follows. In the next section we describe the 3HDM scalar potential in general, followed by  a particular focus on the CPV I(2+1)HDM case. After this, we will define BPs amenable to experimental investigation compliant with current theoretical and experimental limits, the latter crucially including those from both the active Higgs sector and the inert one, as extracted from both DM and collider facilities. We then proceed to describe the physical process targeted at  a future electron-positron collider. After this, we will describe our key observables, i.e., triple products, which represent an ideal context where to pursue evidence of CP-violation. Our results and conclusions will follow in turn.

\section{I(2+1)HDM construction}

In this section, we describe the potential of the $Z_2$-symmetric I(2+1)HDM. As a first step, the doublets $\phi_1$,$\phi_2$ and $\phi_3$ transform as follows:
\begin{eqnarray}\label{config}
    \phi_1\rightarrow -\phi_1,\ \phi_2\rightarrow -\phi_2,\ \phi_3\rightarrow \phi_3.
\end{eqnarray}

The potential consists of two distinct parts: $V_0$, which is invariant under any phase rotation, and $V_{Z_2}$, which is invariant under the $Z_2$ symmetry configuration presented in Eq. (\ref{config}). In compact notation, they are:

\begin{eqnarray}
V_0& = &\sum_{i=1}^3 \left(-\mu_i^2(\phi_i^\dagger\phi_i)+\lambda_{ii}(\phi_i^\dagger\phi_i)^2 \right. +\lambda_{i(i+1)}(\phi_i^\dagger\phi_i)(\phi_{i+1}^\dagger\phi_{i+1})\nonumber\\
&& \left. + \lambda'_{i(i+1)}(\phi_i^\dagger\phi_{i+1})(\phi_{i+1}^\dagger\phi_i)\right),\label{v0}\\
V_{Z_2}& = & -\mu_{12}^2(\phi_1^\dagger\phi_2) + \sum_{i=1}^3\lambda_i(\phi_i^\dagger\phi_{i+1})^2  + \text{h.c.}, \label{vz2}
\end{eqnarray}
where the index sum $i+1$ is a module $3$ term. The parameters in $V_0$  are all real by construction, but in $V_{Z_2}$
 we allow for complex ones, leading to explicit CP-violation in the scalar sector.

The field content of the model includes three doublets, which are defined as follows:
\begin{equation}\label{eq:phis}
\phi_1 = \begin{pmatrix} H_1^+ \\ \frac{H_1^0 + iA_1^0}{\sqrt{2}} \end{pmatrix}, \quad 
\phi_2 = \begin{pmatrix} H_2^+ \\ \frac{H_2^0 + iA_2^0}{\sqrt{2}} \end{pmatrix}, \quad 
\phi_3 = \begin{pmatrix} G^+ \\ \frac{v + h + iG^0}{\sqrt{2}}. \end{pmatrix}
\end{equation}
In this setup, $\phi_1$ and $\phi_2$ are designated as \textit{inert} doublets, being odd under the $Z_2$ discrete symmetry. In contrast, $\phi_3$ acts as the \textit{active} doublet (even under $Z_2$), taking on the role of the SM Higgs doublet. Here, $H^0_i, A^0_i$ and $H^\pm_i$ are the neutral and charged inert Higgs state whereas 
$h$ represents the physical SM Higgs boson, with $G^0$ and $G^\pm$ denoting the associated Goldstone bosons.

The Yukawa Lagrangian for this model is identical in form to the SM one, where only the active doublet $\phi_3$ couples to the fermions:
\begin{equation}
\mathcal{L}_{\text{Yukawa}} = \Gamma^u_{mn} \bar{q}_{m,L} \tilde{\phi}_3 u_{n,R} + \Gamma^d_{mn} \bar{q}_{m,L} \phi_3 d_{n,R} + \Gamma^e_{mn} \bar{l}_{m,L} \phi_3 e_{n,R} + \Gamma^\nu_{mn} \bar{l}_{m,L} \tilde{\phi}_3 \nu_{n,R} + \text{h.c.}
\end{equation}
The terms $\Gamma^{u,d,e,\nu}_{mn}$ are the dimensionless Yukawa couplings for the family indices $m, n$, following the standard notation for SM fermions. 
It is evident that the vacuum alignment $(0, 0, v/\sqrt{2})$ preserves the symmetry of both the scalar potential and Yukawa sector.

As previously mentioned, the scalar $h$ in $\phi_3$ (the unique $Z_2$-even active doublet) is identified with the SM Higgs boson and its couplings are identical to those of the latter. Consequently, CP-violation is introduced only via the inert scalar sector. The DM candidate is thus the lightest mixed CP-eigenstate from the inert scaler sector. To stabilize the DM candidate against decay into SM particles, we exploit the remnant symmetry of the potential following Electro-Weak Symmetry Breaking
(EWSB) \cite{Ivanov:2012hc}.

In the remainder of this work, we will study the dark democracy limit, which is a simplified version of the I(2+1)HDM obtained from the following conditions:
\begin{equation}\label{condi}
\mu_1^2=\mu_2^2,\ \lambda_3=\lambda_2,\ \lambda_{31}=\lambda_{23},\ \lambda_{31}'=\lambda_{23}'.
\end{equation}
As mentioned in several references, see, e.g., \cite{Haber:2006ue,Haber:2015pua,Cordero-Cid:2016krd}, explicit CP-violation is still present after imposing the dark democracy limit. Once the limit is applied, only two parameters remain complex. These are:
\begin{eqnarray}
    \mu_{12}^2&=&\text{Re}(\mu_{12})^2+i \text{Im}(\mu_{12}^2)=|\mu_{12}^2|e^{i\theta_{12}}\label{CPVequation}\\
    \lambda_2&=&\text{Re}(\lambda_{2})+i \text{Im}(\lambda_{2})=|\lambda_{2}|e^{i\theta_{2}}.
\end{eqnarray}
Therefore, the angles $\theta_{12}$ and $\theta_2$ are the CPV phases. 

Since only the $\phi_3$ doublet is active, the minimum of the potential is located at $(0, 0, v/\sqrt{2})$, where $v$ is the SM VEV. The minimization condition requires that $v^2 =\mu_3^2/\lambda_{33}$. Consequently, the mass spectrum for the active doublet is identical to that of the SM, consisting of the Goldstone bosons ($G^0, G^\pm$) and the SM Higgs state ($h$), see Eq.~(\ref{eq:phis}). Furthermore, the inert sector yields two physical charged states ($S_1^\pm$ and $S_2^\pm$) and four neutral CP-mixed ones ($S_1, S_2, S_3, S_4$). The expressions listed below present the relationship between  gauge and mass eigenstates (for more details, see Ref.~\cite{Cordero-Cid:2016krd}):
\begin{eqnarray}
 H_1^\pm &=& \frac{e^{\pm i\theta_{12}/2}}{\sqrt{2}}(S_1^\pm - S_2^\pm),\\ 
 H_2^\pm &=& \frac{e^{\mp i\theta_{12}/2}}{\sqrt{2}}(S_1^\pm + S_2^\pm),\\
S_1 &=& \frac{\alpha H_1^0 + \alpha H_2^0 - A_1^0 + A_2^0}{\sqrt{2\alpha^2 + 2}},\\
S_2 &=& \frac{-H_1^0 - H_2^0 - \alpha A_1^0 + \alpha A_2^0}{\sqrt{2\alpha^2 + 2}}, \\
S_3 &=& \frac{\beta H_1^0 - \beta H_2^0 + A_1^0 + A_2^0}{\sqrt{2\beta^2 + 2}}, \\
S_4 &=& \frac{-H_1^0 + H_2^0 + \beta A_1^0 + \beta A_2^0}{\sqrt{2\beta^2 + 2}},
\end{eqnarray}
where we have used the following definitions: 
\begin{eqnarray}
\Lambda &=& \sqrt{v^4|\lambda_2|^2 + |\mu_{12}^2|^2 - 2v^2|\lambda_2||\mu_{12}^2|\cos(\theta_{12} + \theta_2)}, \label{masa1}\\
\Lambda' &=& \sqrt{v^4|\lambda_2|^2 + |\mu_{12}^2|^2 + 2v^2|\lambda_2||\mu_{12}^2|\cos(\theta_{12} + \theta_2)}, \\
\alpha &=& \frac{-|\mu_{12}^2|\cos \theta_{12} + v^2|\lambda_2|\cos \theta_2 - \Lambda}{|\mu_{12}^2|\sin \theta_{12} + v^2|\lambda_2|\sin \theta_2},\\
\beta &=& \frac{|\mu_{12}^2|\cos \theta_{12} + v^2|\lambda_2|\cos \theta_2 - \Lambda'}{|\mu_{12}^2|\sin \theta_{12} - v^2|\lambda_2|\sin \theta_2}.\label{masa4}
\end{eqnarray}

In general, the mass spectrum is described as follows: 
\begin{eqnarray}
m_{S_1}^2 &=& \frac{v^2}{2}(\lambda'_{23} + \lambda_{23}) - \Lambda - \mu_2^2, \\
m_{S_2}^2 &=& \frac{v^2}{2}(\lambda'_{23} + \lambda_{23}) + \Lambda - \mu_2^2, \\
m_{S_3}^2 &=& \frac{v^2}{2}(\lambda'_{23} + \lambda_{23}) - \Lambda' - \mu_2^2, \\
m_{S_4}^2 &=& \frac{v^2}{2}(\lambda'_{23} + \lambda_{23}) + \Lambda' - \mu_2^2.
\end{eqnarray}
 Essentially, any scalar can be the lightest particle and therefore the DM candidate, however, without any loss of generality, we take $S_1$
as our lightest state. This choice imposes certain conditions on 
$\Lambda$ and $\Lambda '$: we require these to be positive and $\Lambda'<\Lambda$. From these conditions, it follows that $\pi/2<\theta_{12}+\theta_2<\pi$. 

\subsection{Exact-CPC and CPC-limit configurations}

  Since this paper focuses on identifying observables that allow us to distinguish between CPV and CPC scalar states, this subsection presents a brief description of the 
  Exact-CPC and CPC-limit of the CPV I(2+1)HDM.

The CPC I(2+1)HDM is described by the same potential presented in Eqs. (\ref{v0}) and (\ref{vz2}) , but in this case all parameters are taken to be real. Thus, after EWSB, all the mass eigenstates have a well-defined CP parity: we have one SM-like Higgs boson $h$ (CP-even) two dark CP-even scalar bosons ($H_1$ and $H_2$), two dark CP-odd pseudoscalar bosons ($A_1$ and $A_2$) and four charged dark scalar bosons ($H_1^{\pm}$ and $H_2^\pm$). 

  \begin{table}
    \centering
        \begin{tabular}{|c|c|}
    \hline
      Exact-CPC & CPC-limit  \\
     \hline
     \hline
      $H_1=\frac{H_1^0+H_2^0}{\sqrt{2}},\ m_{H_1}^2=\frac{v^2}{2}(\lambda_{23}+\lambda'_{23}+2\lambda_2)-\mu_2^2-\mu_{12}^2$ & $S_1,\ m_{S_1}^2$  \\
      \hline
       $H_2=\frac{-H_1^0+H_2^0}{\sqrt{2}},\ m_{H_2}^2=\frac{v^2}{2}(\lambda_{23}+\lambda'_{23}+2\lambda_2)-\mu_2^2+\mu_{12}^2$ & $S_3,\ m_{S_3}^2$  \\
      \hline
      $A_1=\frac{A_1^0+A_2^0}{\sqrt{2}},\ m_{A_1}^2=\frac{v^2}{2}(\lambda_{23}+\lambda'_{23}-2\lambda_2)-\mu_2^2-\mu_{12}^2$ & $S_4,\ m_{S_4}^2$  \\
      \hline
       $A_2=\frac{-A_1^0+A_2^0}{\sqrt{2}},\ m_{A_2}^2=\frac{v^2}{2}(\lambda_{23}+\lambda'_{23}-2\lambda_2)-\mu_2^2+\mu_{12}^2$ & $S_2,\ m_{S_2}^2$  \\
       \hline 
       $\textbf{H}_1^{\pm}=\frac{H_1^\pm+H_2^\pm}{\sqrt{2}}$, $M_{H_1^\pm}^2=\frac{\lambda_{23}v^2}{2}-\mu_2^2-\mu_{12}^2$ & $S_1^\pm,\ M_{S_1^\pm}^2$\\
       \hline
        $\textbf{H}_2^{\pm}=\frac{H_1^\pm-H_2^\pm}{\sqrt{2}}$, $M_{H_2^\pm}^2=\frac{\lambda_{23}v^2}{2}-\mu_2^2+\mu_{12}^2$ & $S_2^\pm,\ M_{S_2^\pm}^2$\\
       \hline
    \end{tabular}
    \caption{Relationship (and definitions) between Exact-CPC fields and CPV-limit ones. }
    \label{masas}
\end{table}

Table~\ref{masas} shows the relation between the Exact-CPC fields and their CPC-limit counterparts, where the same conditions as in  Eq.~(\ref{condi}) are applied (see Ref.~\cite{Keus:2014jha} for details). The mass relations are obtained in the limit $\theta_2+\theta_{12}\rightarrow \pi$ (see Eqs.~(\ref{masa1})--(\ref{masa4})). However, to satisfy the definitions of the mass eigenstates, $\alpha$ and $\beta$ must approach infinity, which occurs when $\theta_2\rightarrow \pi$ and $\theta_{12}\rightarrow 0$. In fact, when $\cos(\theta_2+\theta_{12})=-1$,
we obtain
  \begin{eqnarray}
      \Lambda&=&v^2|\lambda_2|+|\mu_{12}^2|,\\
      {
      \Lambda'
      }
      &=&v^2|\lambda_2|-|\mu_{12}^2|,
  \end{eqnarray} 
Furthermore, the coupling $g_{h\,{\rm DM\,DM}}$ of the CPC case is achieved when $\alpha$ tends to infinity  \cite{Keus:2014jha}, 
 \begin{eqnarray}
     g_{h\,{\rm DM\,DM}}^{\rm CPC} &=&\lim_{\alpha\to\infty}g_{h\,{\rm DM\,DM}}^{\rm CPV}\nonumber\\
     &=& \lim_{\alpha\to\infty}\frac{1}{1 + \alpha^2} \left[ 4\alpha \text{Im}\lambda_2 + 2(\alpha^2 - 1) \text{Re}\lambda_2 \right] + \lambda_{23} + \lambda'_{23} \\
&=& 2\lambda_2 + \lambda_{23} + \lambda'_{23},
 \end{eqnarray}

{
For the primary purpose of this work, another coupling of utmost importance is the one related to the $Z\phi_i \phi_j$ interaction. 
In the CPC case, this coupling is exclusively possible for $ZH_1A_1$. 
In contrast, in the CPV case, the undefined parity states allow the couplings $ZS_1S_3$ and $ZS_1S_4$ to exist.
}
These have the following form:
\begin{eqnarray}
    g_{ZS_1S_3}&\propto& \frac{\alpha+\beta}{\sqrt{\alpha^2+1}\sqrt{\beta^2+1}},\label{s2s3} \\
    g_{ZS_1S_4}&\propto&\frac{\alpha\beta-1}{\sqrt{\alpha^2+1}\sqrt{\beta^2+1}}.\label{s2s4}
\end{eqnarray}
Once again, in the limit where $\alpha$ approaches {infinity} (as well as $\beta$), the coupling $g_{ZS_1S_3}$ approaches zero, whereas the coupling $g_{ZS_1S_4}$ tends to the explicit CPC form.
In fact, the Lagrangian term describing this coupling can be written in a general form as follows
\begin{equation}
\mathcal{L}_{Z S_l S_r} = \frac{g}{2\sqrt{2} c_W} \left[ (q_{1r} q_{2l} - q_{1l} q_{2r}) + (q_{3r} q_{4l} - q_{4r} q_{3l}) \right] \times Z^\mu S_l{\partial}_\mu S_r 
\end{equation}
with the $q$'s represent the entries of the mixing matrix that rotates the gauge eigenstates into the mass eigenstates
\begin{equation}
    \begin{pmatrix}
H_1 \\
A_1 \\
H_2 \\
A_2
\end{pmatrix}
= \frac{1}{\sqrt{2}}\begin{pmatrix}
\frac{\alpha}{\sqrt{1+\alpha^2}} & -\frac{1}{\sqrt{1+\alpha^2}} & \frac{\beta}{\sqrt{1+\beta^2}} & -\frac{1}{\sqrt{1+\beta^2}} \\
\frac{\alpha}{\sqrt{1+\alpha^2}} & -\frac{1}{\sqrt{1+\alpha^2}} & -\frac{\beta}{\sqrt{1+\beta^2}} & \frac{1}{\sqrt{1+\beta^2}} \\
-\frac{1}{\sqrt{1+\alpha^2}} & -\frac{\alpha}{\sqrt{1+\alpha^2}} & \frac{1}{\sqrt{1+\beta^2}} & \frac{\beta}{\sqrt{1+\beta^2}} \\
\frac{1}{\sqrt{1+\alpha^2}} & \frac{\alpha}{\sqrt{1+\alpha^2}} & \frac{1}{\sqrt{1+\beta^2}} & \frac{\beta}{\sqrt{1+\beta^2}}
\end{pmatrix}
\begin{pmatrix}
S_1 \\
S_2 \\
S_3 \\
S_4
\end{pmatrix}.
\end{equation}
In the specific cases of interest here, setting $r=1$ and $l=3,4$ recovers expresions shown in ecuations \ref{s2s3},\ref{s2s4}, while for $l=2$ it vanishes identically.

\section{Parameter space}

\begin{figure}
     \centering
     \includegraphics[width=0.4\linewidth]{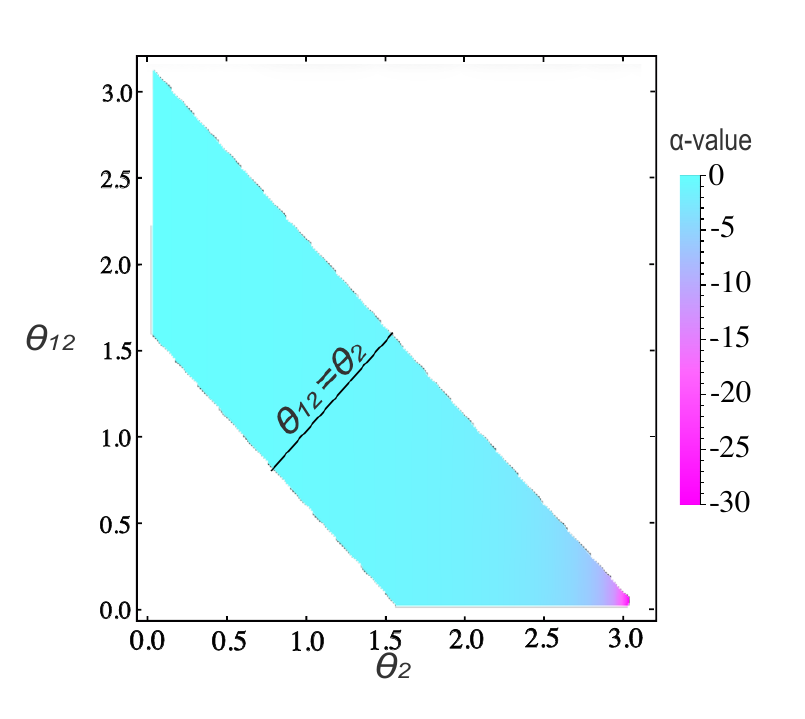}
      \includegraphics[width=0.4\linewidth]{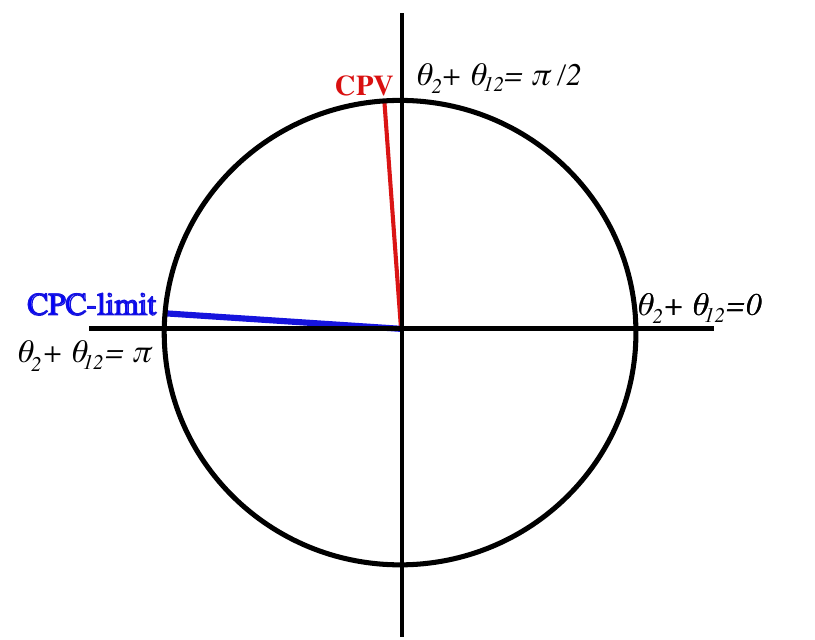}
     \caption{Values of $\alpha$ for possible configurations of $\theta_2$ and $\theta_{12}$. The black line represent the values of $\alpha$ when $\theta_2=\theta_{12}$.} 
     \label{alpha}
 \end{figure}

In previous work, the theoretical and experimental constraints on the model were discussed in Ref.~\cite{Cordero-Cid:2016krd}. In the present paper, we reconsider similar parameter values and update DM observables therein by using those calculated by \texttt{micrOMEGAs 6.2} \cite{Alguero:2023zol}. Here, the 
 task is to analyze observables sensitive to CP-violation, thus, the key parameters for our study are the CPV phases ($\theta_2$ and $\theta_{12}$). To start with, we require $\pi/2<\theta_2+\theta_{12}<\pi$ to ensure that $S_1$ is the lightest field and the DM candidate. 
Figure \ref{alpha} shows the behavior of the $\alpha$ parameter for the accepted values of $\theta_2$ and $\theta_{12}$, assuming that $S_1$ is the lightest scalar field. As it can be observed, it is imperative to take the $\theta_2$  values very close to $\pi$ if one wants to achieve the CPC configuration (henceforth, CPC-limit). However, if $\theta_2$ is very close to $\pi$ (and $\theta_{12}$ is close to zero), significant numerical instabilities can occur in our calculations. To avoid this, we introduce the CPC model 
at Lagrangian level (henceforth, Exact-CPC) for the computation. This approach will be especially important for selecting our BPs.

\begin{figure}
    \centering
    \includegraphics[width=0.7\linewidth]{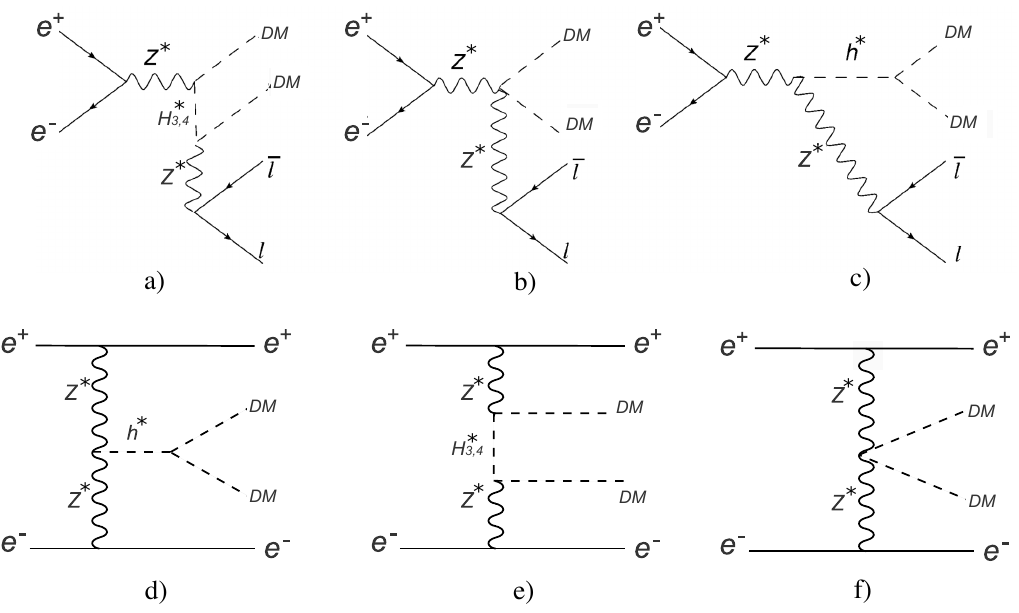}
    \caption{Diagrams for the $e^+e^-\to  l^+l^-$ DM DM process 
     in the I(2+1)HDM with CP-violation, with the DM pair yielding $\cancel{E}_T$. Here, $H_{3,4}$ refer to $S_{3,4}$ in the CPV case, while only to $A_1$ in the CPC one.}
    \label{diag-proce}
\end{figure}

The primary objective of this work is to demonstrate the feasibility of distinguishing between final states associated with the CPV and CPC regimes in experimental searches. We focus on the process $e^+ e^- \to l^+ l^- + \cancel{E}_T$ (with $l^\pm=\{e^\pm,\mu^\pm\}$), %
where the $\cancel{E}_T$ arises from the production of DM  pairs in either the CPV  or CPC case (see the diagrams in Figure~\ref{diag-proce}).

{
Based on the considerations detailed in this section, we define two BPs, which  parameter values are summarized in Table~\ref{tab:benchmark_points}. The first one, BP-CPC, is an Exact-CPC scenario. As discussed above, reaching the CPC-limit from within the CPV model would require $\theta_2\to\pi$ and $\theta_{12}\to 0$, that is $\alpha,\beta\gg 1$, a region in which the numerical evaluation of the spectrum becomes unstable.  As mentioned, we therefore use the Exact-CPC model as our CPC reference rather than a BP sitting in that limit. The second one, BP-CPV, covers the CPV regime and has been chosen close to the extreme CPV configuration $\theta_2+\theta_{12}\approx\pi/2$. The two points share the same DM mass, the same mass splittings as well as the same value of $g_{h\,{\rm DM\,DM}}$, so that any difference between the corresponding distributions would come from the CPV phases alone.
}

\begin{table}[ht]
\centering
\begin{tabular}{lccccccccccc}
\hline \hline
Case & $\theta_{12}$ & $\theta_2$ & $g_{h\,{\rm DM\,DM}}$ & $m_{S1}$ & $\delta_{1s}$ & $\delta_{1c}$ & $\delta_c$ & $\Omega_{\rm DM}h^2$&$\sigma_{\rm SI}$&$\langle\sigma v\rangle$ \\ \hline
BP-CPC & --- & --- & 0.002 & 62.0 & 125.0 & 50.0 & 50.0&$3.06\times10^{-3}$&$1.79\times 19^{-47}$&$2.35\times 10^{-27}$ \\
BP-CPV & 0.02 & 1.57 & 0.002 & 62.0 & 125.0 & 50.0 & 50.0 &$2.88\times10^{-3}$&$1.79\times 10^{-47}$&$2.29\times 10^{-27}$ \\
\hline \hline
\end{tabular}
\caption{Parameter values and main experimental constraints for the selected BPs in the CPC and CPV regimes. Masses and splitting parameters are given in GeV. Here, the splitting parameters are defined as $\delta_{1s}=m_{s_2}-m_{s_1}$, $\delta_{1c}=m_{s_1^\pm}-m_{s_1}$ and $\delta_c=m_{s_2^\pm}-m_{s_1^\pm}$. The Spin-Independent DM-nucleon cross section ($\sigma_{\rm SI}$) is presented in $cm^2$ and the thermally averaged one ($\langle\sigma v\rangle$) is presented in $cm^{3}s^{-1}$.
{The relic density obtained for both BPs lies about a factor of $40$ below the measured value $\Omega_{\rm DM}h^2=0.120$, as expected for a DM mass located in the $h$-funnel ($m_{S_1}\simeq m_h/2$). The lightest inert state therefore accounts only for a subdominant fraction $\xi=\Omega_{\rm DM}h^2/0.120\simeq 2.5\times 10^{-2}$ of the observed DM and the corresponding rate has to be compared with experiment after rescaling by $\xi$. With this rescaling both BPs are compatible with all present direct and indirect detection bounds.}}
\label{tab:benchmark_points}
\end{table}

\section{Process choice}

In this section, we discuss the characteristics of the aforementioned $e^+e^-$ process. Within the CPV I(2+1)HDM, this process proceeds via the Feynman diagrams depicted in Figure \ref{diag-proce} (analogous diagrams also contribute to the CPC case). This graph can be categorized into two main topologies: the three diagrams in the upper panel describe $s$-channel interactions, where the final-state leptons (either electrons or muons) are produced secondarily. Conversely, the three diagrams in the lower panel are exclusive to final states with electrons, where the DM candidates are produced via $t$-channel exchange.

The two general channels discussed above exhibit significant differences from each other. Table~\ref{S-tchannels} presents several physical observables that highlight these distinctions. From the table, it is clear that the $t$-channel contribution is 
{always subleading with respect to the $s$-channel one. The size of the hierarchy, however, depends strongly on the collision energy: it goes from three orders of magnitude at $\sqrt{s}=250$ GeV down to roughly a factor of two at $\sqrt{s}=1000$ GeV. Moreover, at all the energies considered, the total cross section is smaller than the sum of the two exclusive contributions, i.e., the two topologies interfere destructively.}

As will be shown in the Results section, certain observables are highly sensitive to both topologies. In particular, the pseudorapidity of the final-state leptons ($\eta (l)$) and their angular separation in the $\eta-\phi$ plane ($\Delta R(l^+,l^-)$) represent the most promising observables for distinguishing between these two contributions.

\begin{table}
    \centering
    \begin{tabular}{cccc}
       \hline\hline
        BP & $\sigma_{250} $ &$\sigma_{500}$&$\sigma_{1000}$ \\
       \hline\hline
        BP-CPC & $2.89$ ($2.93$) [$0.002$] &$1.14$ ($1.14$) [$0.027$]  & $0.21$ ($0.19$) [$0.083$]\\
        BP-CPV & $0.08$ ($0.095$) [$0.003$] & $1.81$ ($1.83$) [$0.036$]&$0.74$ ($0.67$) [$0.117$] \\
        \hline\hline
    \end{tabular}
    \caption{
    {
    Cross sections in fb for the two BPs at $\sqrt{s}=250$, $500$ and $1000$ GeV.
    } 
    Plain values represent the total cross sections while values in round  (...) and square brackets [...] denote the exclusive $s$-channel and $t$-channel contributions, respectively.
    }
    \label{S-tchannels}
\end{table}

Regarding the diagrams shown in Figure~\ref{diag-proce}, the same topologies are present for both the CPC and CPV cases. However, due to the CP-mixing states in the CPV scenario, the diagrams labeled as \textbf{a)} and \textbf{e)} receive contributions from the $S_3$  and $S_4$ scalar fields whereas, in the CPC scenario, only the $A_1$ contribution is allowed.

\section{Triple Product Observables}

In high-energy $e^+ e^-$ collision processes, kinematic triple vector products constructed from initial-state momenta, final-state momenta and the $\cancel{E}_T$ vector (hereafter, $\mathbf{MET}$)  serve as sensitive probes for CPV structures and kinematic asymmetries. Considering the momentum vectors of the initial- and 
final-state electron-positron pairs alongside $\mathbf{MET}$, three primary triple product configurations can be formulated:
\begin{itemize}
    \item \textbf{Initial-state momenta and MET:}
    \begin{equation}
        \mathbf{MET} \cdot (\mathbf{P}_{i e^-} \times \mathbf{P}_{i e^+}).
    \end{equation}
    
    \item \textbf{Final-state momenta and MET:}
    \begin{equation}
        \mathbf{MET} \cdot (\mathbf{P}_{f e^-} \times \mathbf{P}_{f e^+}).
    \end{equation}
    \item \textbf{Mixed initial-final state momentum and MET:}
    \begin{equation}
        \mathbf{MET} \cdot (\mathbf{P}_{i e^-} \times \mathbf{P}_{f e^-}).
    \end{equation}
\end{itemize}

The first triple product configuration vanishes identically, $\mathbf{MET} \cdot (\mathbf{P}_{i e^-} \times \mathbf{P}_{i e^+}) = 0$, due to the exact collinearity of the counter-propagating initial-state beams along the longitudinal $z$-axis.
For the second configuration involving two final-state momenta, we recall the definition of the missing transverse momentum vector in the transverse plane:
\begin{equation}
    \mathbf{MET} = -\left(P_{f e^-}^x + P_{f e^+}^x, \, P_{f e^-}^y + P_{f e^+}^y, \, 0\right).
\end{equation}
Using Levi-Civita index notation, this triple scalar product is expressed as:
\begin{align}
    \mathbf{MET} \cdot (\mathbf{P}_{f e^-} \times \mathbf{P}_{f e^+}) &= \epsilon_{ijk} \text{MET}^i P_{f e^-}^j P_{f e^+}^k \nonumber \\
    &= -\epsilon_{ijk} \left(P_{f e^-}^i + P_{f e^+}^i\right) P_{f e^-}^j P_{f e^+}^k \nonumber \\
    &= -\epsilon_{ijk} P_{f e^-}^i P_{f e^-}^j P_{f e^+}^k - \epsilon_{ijk} P_{f e^+}^i P_{f e^-}^j P_{f e^+}^k \nonumber \\
    &= -\epsilon_{xjk} P_{f e^-}^x P_{f e^-}^j P_{f e^+}^k - \epsilon_{xjk} P_{f e^+}^x P_{f e^-}^j P_{f e^+}^k \nonumber \\
    &\quad - \epsilon_{yjk} P_{f e^-}^y P_{f e^-}^j P_{f e^+}^k - \epsilon_{yjk} P_{f e^+}^y P_{f e^-}^j P_{f e^+}^k.
\end{align}
Because $\text{MET}^z = 0$, non-vanishing terms restrict the index $i$ to the transverse plane indices $i \in \{x, y\}$. Furthermore, structural cancellations occur due to symmetry properties: terms with $k = z$ cancel pairwise between the first and third summands, while terms with $j = z$ cancel between the second and fourth summands. Simplifying the remaining non-zero components yields:
\begin{align}
    \mathbf{MET} \cdot (\mathbf{P}_{f e^-} \times \mathbf{P}_{f e^+}) &= -\epsilon_{xzy} P_{f e^-}^x P_{f e^-}^z P_{f e^+}^y - \epsilon_{xyz} P_{f e^+}^x P_{f e^-}^y P_{f e^+}^z \nonumber \\
    &\quad - \epsilon_{yzx} P_{f e^-}^y P_{f e^-}^z P_{f e^+}^x - \epsilon_{yxz} P_{f e^+}^y P_{f e^-}^x P_{f e^+}^z \nonumber \\
    &= \left(P_{f e^-}^z + P_{f e^+}^z\right) \left(P_{f e^-}^x P_{f e^+}^y - P_{f e^+}^x P_{f e^-}^y\right).
\end{align}
All remaining permutations of these three vectors are directly connected through cyclic identities of the scalar triple product.
Finally, for the third configuration involving mixed initial- and final-state momenta alongside $\mathbf{MET}$, the product evaluates to:
\begin{equation}
    \mathbf{MET} \cdot (\mathbf{P}_{i e^-} \times \mathbf{P}_{f e^-}) = -\epsilon_{ijk} \left(P_{f e^-}^i + P_{f e^+}^i\right) P_{i e^-}^j P_{f e^-}^k.
\end{equation}
Given that the initial electron momentum is aligned purely along the beam axis, $\mathbf{P}_{i e^-} = \left(0, 0, \frac{\sqrt{s}}{2}\right)$, only the longitudinal component $j = z$ contributes:
\begin{align}
    \mathbf{MET} \cdot (\mathbf{P}_{i e^-} \times \mathbf{P}_{f e^-}) &= -\epsilon_{izk} \left(P_{f e^-}^i + P_{f e^+}^i\right) P_{i e^-}^z P_{f e^-}^k \nonumber \\
    &= -\epsilon_{xzy} \left(P_{f e^-}^x + P_{f e^+}^x\right) P_{i e^-}^z P_{f e^-}^y - \epsilon_{yzx} \left(P_{f e^-}^y + P_{f e^+}^y\right) P_{i e^-}^z P_{f e^-}^x \nonumber \\
    &= \frac{\sqrt{s}}{2} \left(P_{f e^+}^x P_{f e^-}^y - P_{f e^+}^y P_{f e^-}^x\right).
\end{align}
Similarly, alternative orderings of this triad are linked via cyclic permutation properties.

In summary, 
{
these last two are the only independent functional forms that survive for the process at hand. It is worth stressing, however, that both of them are proportional to the same transverse antisymmetric combination
\begin{equation}
    X \equiv P_{f e^-}^x P_{f e^+}^y - P_{f e^+}^x P_{f e^-}^y ,
\end{equation}
the mixed configuration thus being $X$ multiplied by the constant $-\sqrt{s}/2$, so that it is nothing but $X$ itself up to normalization, with the purely final-state configuration being $X$ weighted by the longitudinal momentum of the lepton pair, $P_{f e^-}^z + P_{f e^+}^z$. Since such a weight is not sign-definite at a symmetric $e^+e^-$ collider, the two observables are not redundant: i.e., an asymmetry built from the sign of the mixed configuration measures the sign of $X$ alone whereas an asymmetry built from the sign of the purely final-state one measures the correlation between $X$ and the longitudinal boost of the pair.
The subsequent figures will display the differential distributions and kinematic histograms obtained for the BPs defined in Table~\ref{tab:benchmark_points}.
}

\section{Results}\label{sec:results}

The following analysis was performed using CalcHEP 3.9.2 \cite{Belyaev:2012qa} as the event generator. Subsequently, MadAnalysis 5 was employed in expert mode to analyze the simulated events \cite{Conte:2012fm}. For each benchmark scenario, $10^5$  events were generated to ensure sufficient statistical precision. All results presented in this work are computed at the parton level.

{Below, we present the numerical results for the various kinematic variables that characterize the $e^+e^-\to l^+l^- +\cancel{E}_T$ process. In this analysis, the DM  candidate is examined within the two limiting cases permitted by the model, namely, the CPC and CPV one, realized through the diagrams of Figure~\ref{diag-proce}.}

This set of BPs presents {cross sections of comparable size at $\sqrt{s}=500$ and $1000$ GeV, whereas at $\sqrt{s}=250$ GeV the CPV BP is suppressed by more than an order of magnitude with respect to the CPC one,} when considering all diagrams depicted in {Figure~\ref{diag-proce}} (see Table~{\ref{S-tchannels}}). However, the final leptons in the lower-panel diagrams of {Figure~\ref{diag-proce}} are expected to be highly forward (or backward), which causes their contribution to the triple product to act as a, so to say, intrinsic background. In fact, the most relevant region in pseudorapidity space for the triple product is where the lepton momentum is highly central (where better detector performance occurs), which {suppresses} the $t$-channel contributions. In this sense, Table~{\ref{S-tchannels}} also presents the cross sections for the $s$- and $t$-channel contributions.

\begin{figure}
    \centering
    \includegraphics[width=0.39\linewidth]{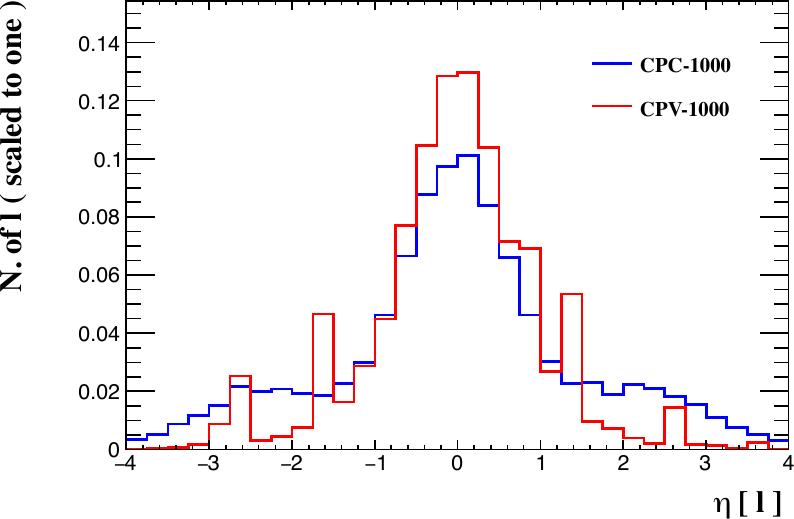}
     \includegraphics[width=0.38\linewidth]{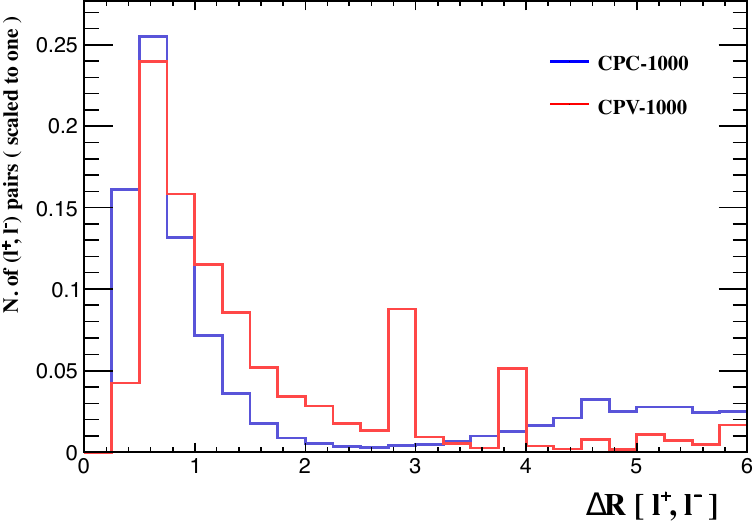}\\
      \includegraphics[width=0.39\linewidth]{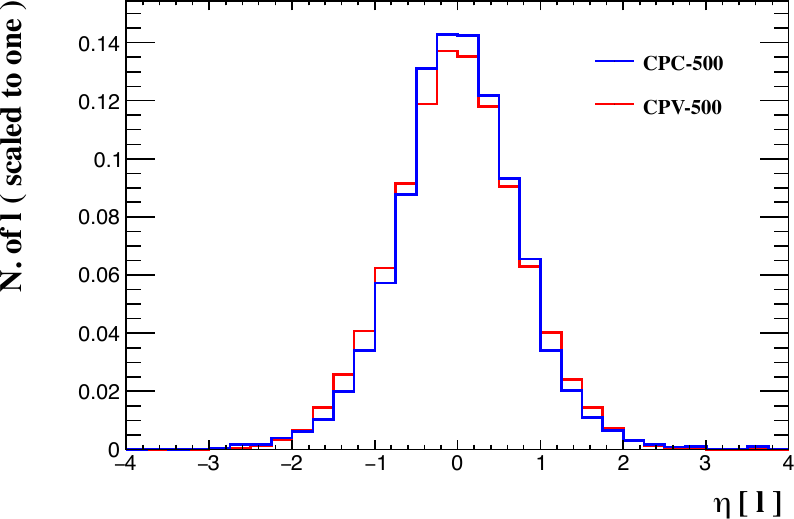}
     \includegraphics[width=0.38\linewidth]{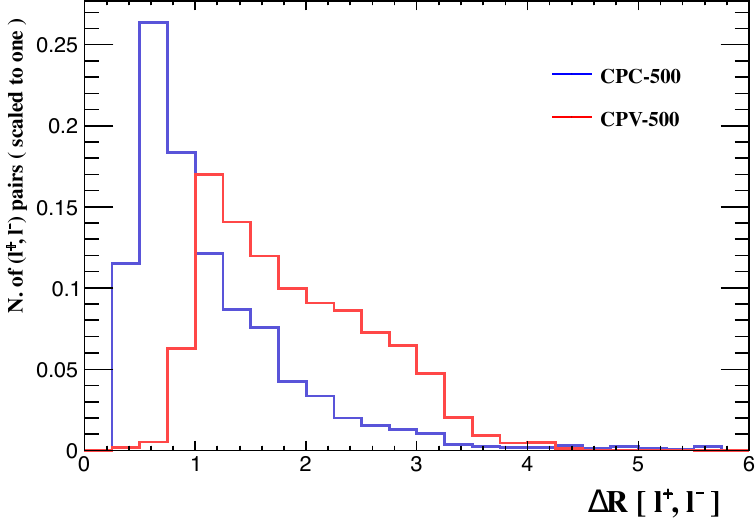}\\
 \includegraphics[width=0.39\linewidth]{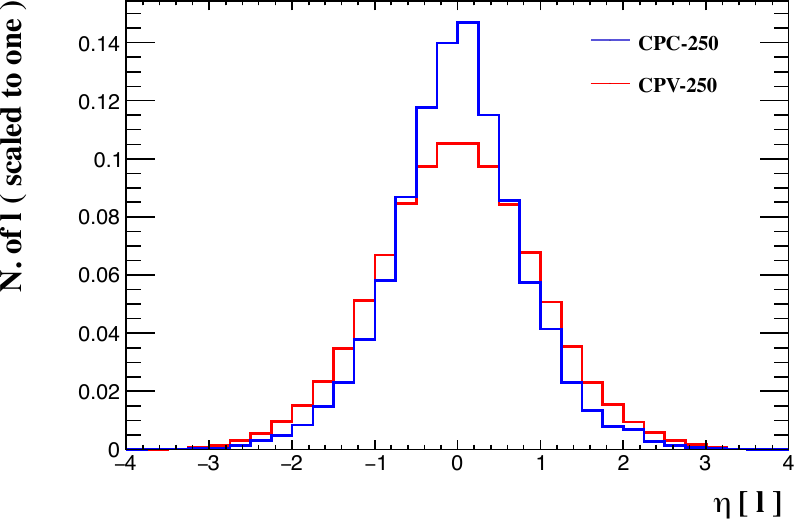}
     \includegraphics[width=0.38\linewidth]{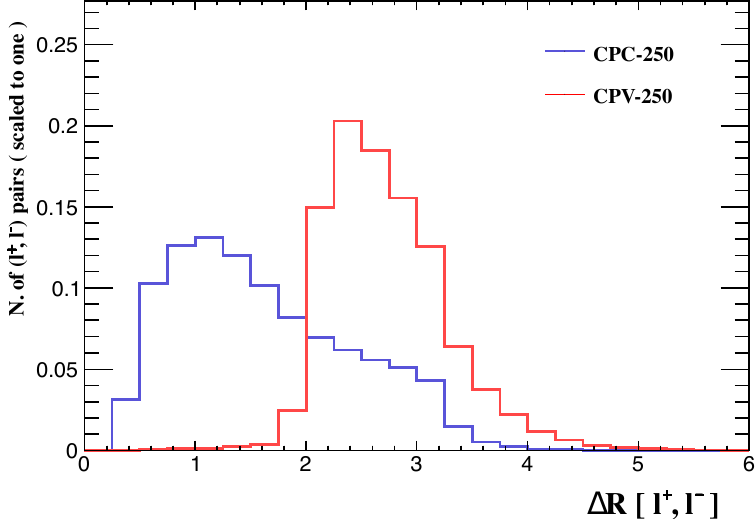}\\

    \caption{{Distributions of the final-state lepton pseudorapidity $\eta(l)$ (left column) and  the angular separation between them in the $\eta-\phi$ plane $\Delta R(l^+,l^-)$ (right column), for $\sqrt{s}=1000$, $500$ and $250$ GeV (from top to bottom). The red(blue) line represents the CPV(CPC) case.}
    }
    \label{observables}
\end{figure}

We begin our analysis by presenting and discussing some physical observables associated 
with the two final-state leptons: the pseudorapidity of each lepton and the relative distance 
between the two leptons in the $\eta-\phi$ plane (where $\phi$ is the azimuthal angle 
of the momentum).  The behavior of these three observables is shown in Figure~\ref{observables}. 
The plots in the left panels make it evident that CPC and CPV final states exhibit a 
distinctly central behavior, with moderate differences. In contrast, the plots in the right panels show 
that, for the CPC setup, the final leptons are more collimated than in the CPV scenario, the more so the lower the collider energy, indicating a different predominance of of the $s$- and $t$-channel topologies in the two cases~\footnote{Notice that the sensitivity of such kinematical variables to the CP properties of DM in this very same process had already been emphasized in  Ref.~\cite{Dey:2024epo}, albeit in the case of two-component DM emerging from 
the I(2+1)HDM with a $Z_2\times Z_2$ symmetry.}. 

\begin{figure}
    \centering
    \includegraphics[width=0.4\linewidth]{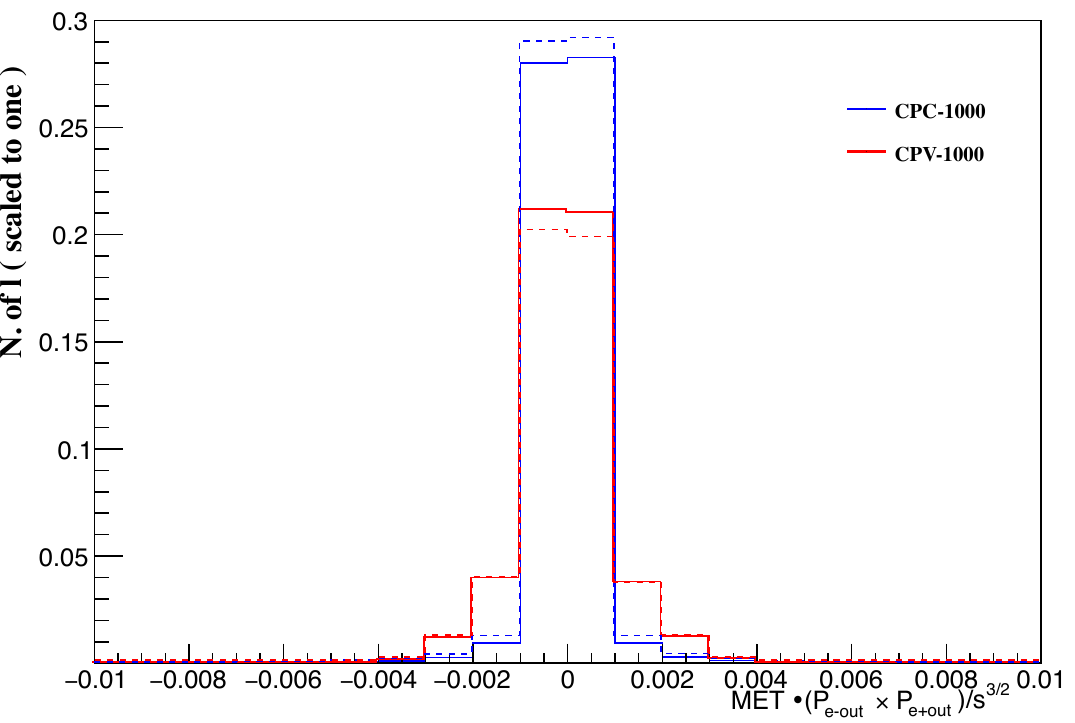}
    \includegraphics[width=0.4\linewidth]{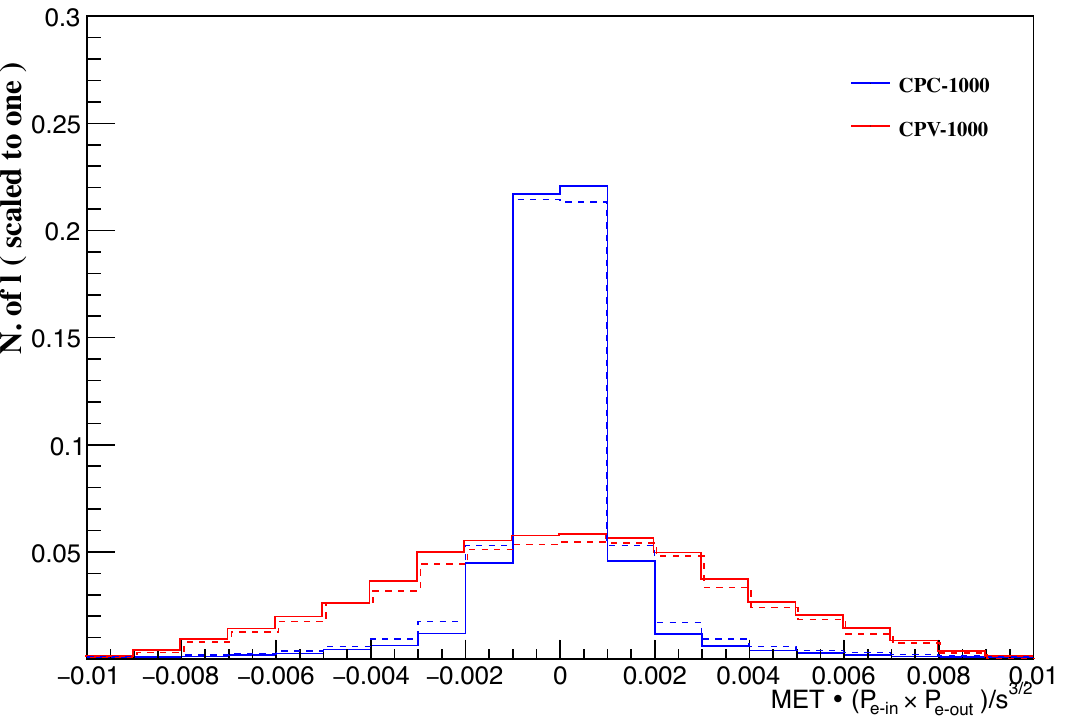}\\
    \includegraphics[width=0.4\linewidth]{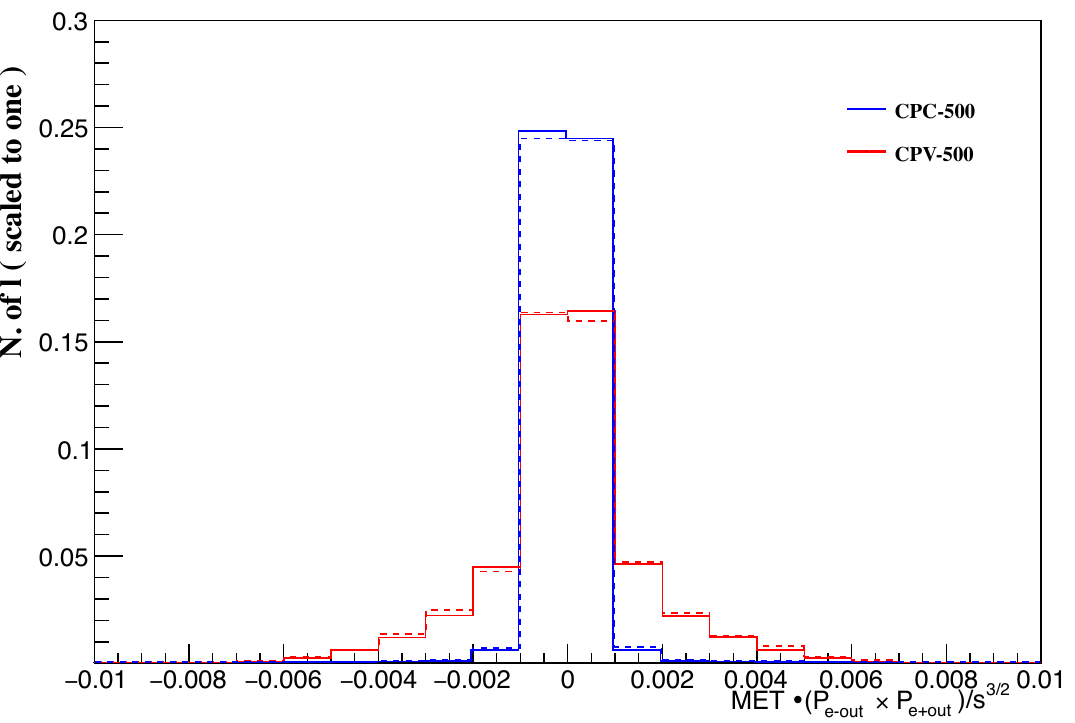}
    \includegraphics[width=0.4\linewidth]{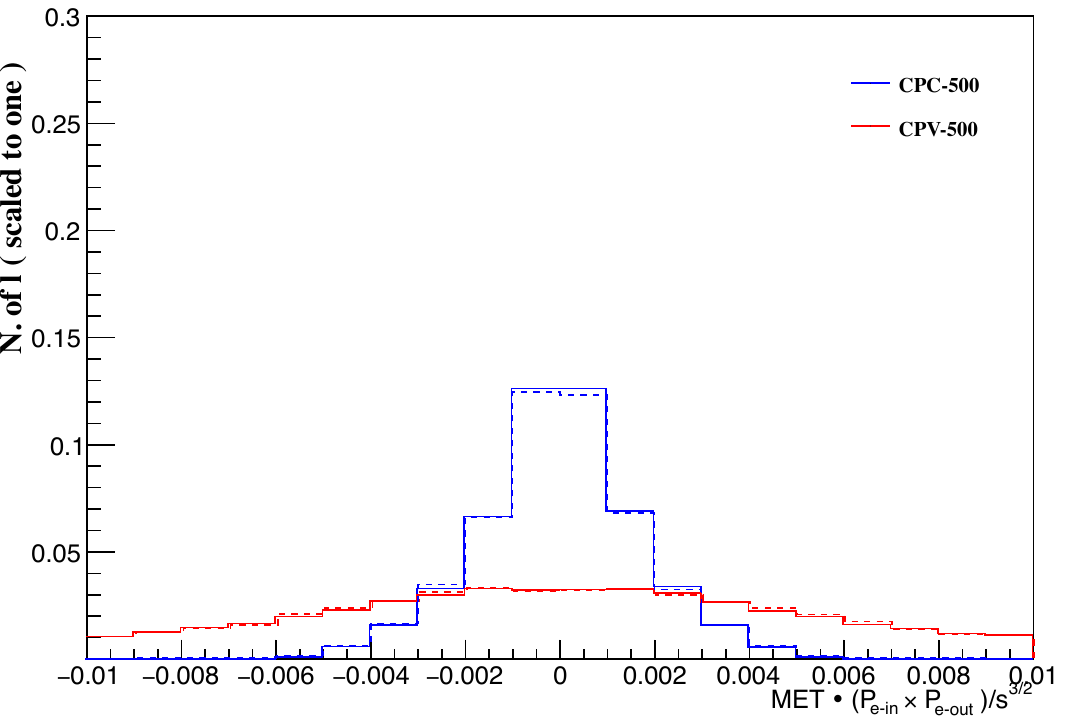}\\
    \includegraphics[width=0.4\linewidth]{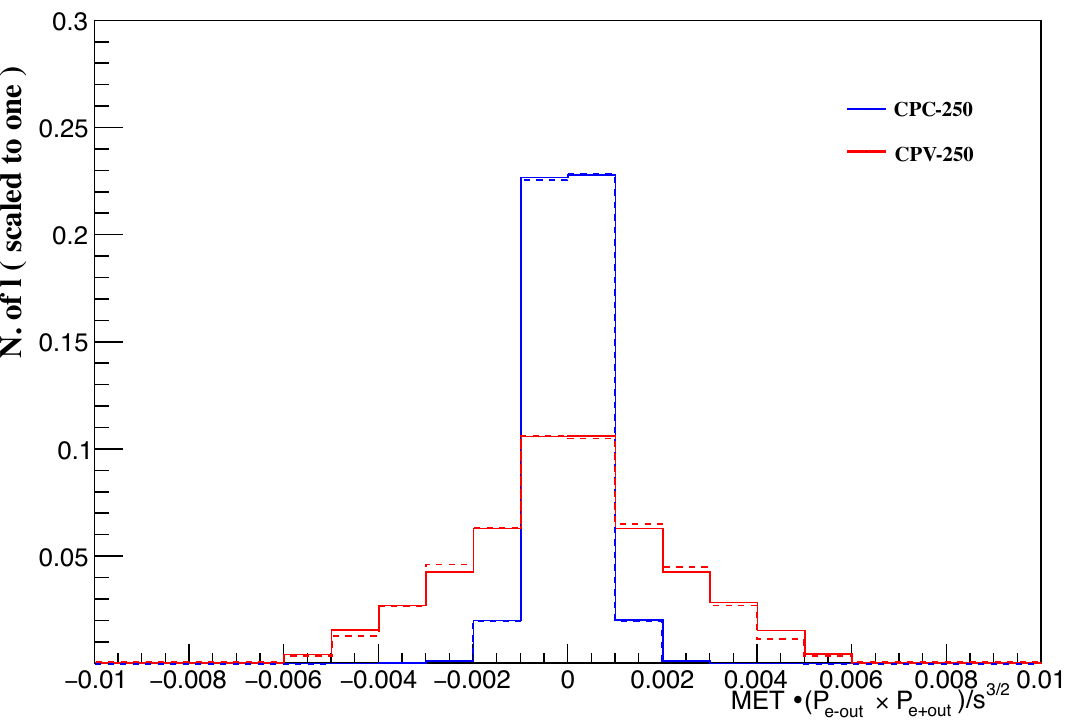}
    \includegraphics[width=0.4\linewidth]{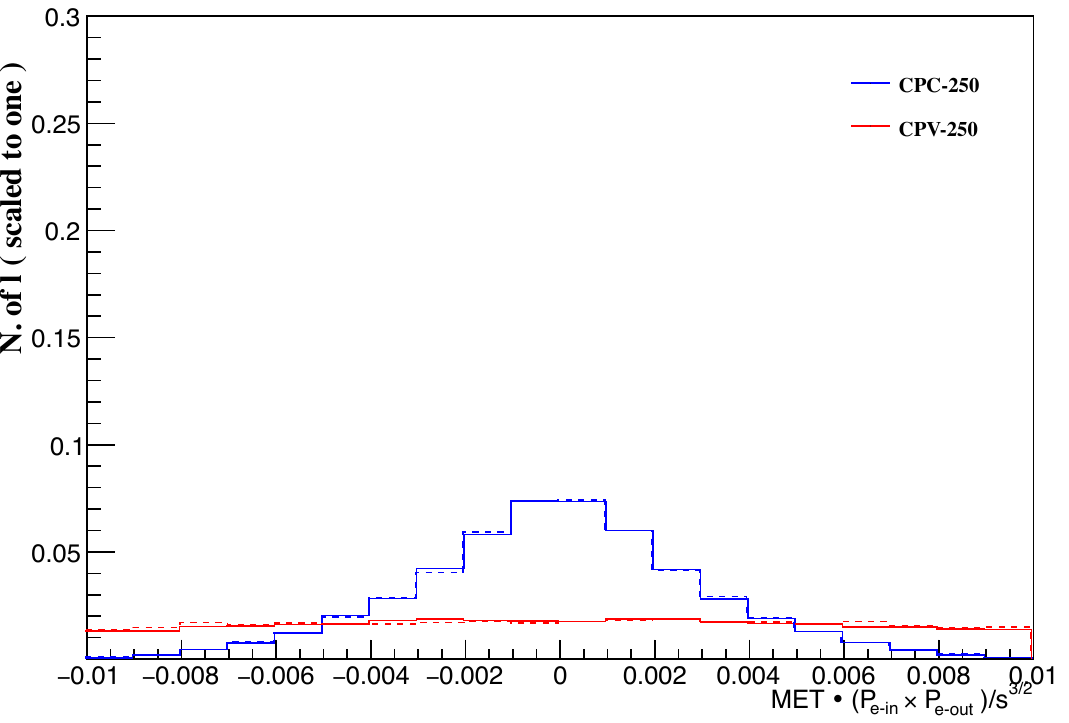}
    \caption{Triple product distributions for the observables   $(\mathbf{MET}\cdot (\mathbf{P}_{fe^-}\times \mathbf{P}_{fe^+}))/{s}^{3/2}$ (left column) and $(\mathbf{MET}\cdot (\mathbf{P}_{ie^-}\times \textbf{P}_{fe^-}))/{s}^{3/2}$ (right column), for $\sqrt{s}=1000$, $500$ and $250$ GeV (from top to bottom). The red(blue) line represents the CPV(CPC) case. The dashed line represents the case with polarized beams (80\% for the electron  and 30\% for the positron).}
    \label{tripleuno}
\end{figure}

As previously emphasized, however, much more sensitive observables for detecting CPV effects are triple products. By adopting the two non-zero ones introduced in the
preceding section, i.e., 
$\mathbf{MET}\cdot(\mathbf{P}_{ie^-}\times \mathbf{P}_{fe^-})$ and $\mathbf{MET}\cdot(\mathbf{P}_{fe^-}\times \mathbf{P}_{fe^+})$. The plots in Figure \ref{tripleuno} display the triple product behavior for two cases: CPV and CPC.
As shown, compatibly with the finite resolution of the detectors on these observables (emulated here through the bin width), that we have normalized here so as to be dimensionless, we notice that {CPC events cluster at zero whereas the CPV ones stretch significantly beyond.} This different behavior relates to the propagator structure of some of the Feynman diagrams intervening in the process we studied, i.e., $S_{3,4}$-mediation in the CPV case vs. $A_{1}$-mediation in the CPC one.
(We also notice that this result is independent of the beam polarization configuration.)

.

\section{Summary and conclusions}

We have studied how explicit CP-violation in the inert sector of the I(2+1)HDM 
with a (single) $Z_2$ symmetry separating all the SM states from the inert ones 
can be exposed at a future $e^+e^-$ collider, with or without beam polarization. 
In the well-established dark democracy limit the potential keeps two physical CPV  phases, $\theta_{12}$ and $\theta_2$, and the requirement that the lightest neutral inert state $S_1$ be the DM candidate restricts these to respect the relation $\pi/2<\theta_{12}+\theta_2<\pi$. 
Within this angular region, all the neutral inert states $S_i$ ($i=1,...4$)
carry no definite CP parity and the gauge interactions of the dark sector are correspondingly modified: both $ZS_1S_3$ and $ZS_1S_4$ vertices are allowed whereas only the second one survives when the CPC-limit $\alpha,\beta\to\infty$ is taken. 
We have thus exploited this property so as to turn it into an observable effect.

We have then selected two BPs lying on the so called $h$-funnel region, wherein  $m_{S_1}\simeq m_h/2$,  which satisfy current DM direct- and indirect-detection limits while providing a subdominant DM component to the relic density and finally  
 computed the process $e^+e^-\to l^+l^-+\cancel{E}_T$ ($l=e,\mu$) for $\sqrt{s}=250$, $500$ and $1000$ GeV, using such BPs as inputs. 
This process receives contributions from $s$-channel and $t$-channel topologies, the latter being present only for electron-positron final states. 
The two interfere destructively at all the energies considered and the relative weight of the $t$-channel grows markedly with the collision energy, from a permille effect at $\sqrt{s}=250$ GeV to a correction of order 50\% at $\sqrt{s}=1000$ GeV.

Among the kinematic variables associated to this process, the pseudorapidity $\eta(l)$ of either of the final-state leptons and their separation $\Delta R(l^+,l^-)$ turn out to be the most useful ones for isolating  CPV effects from the CPC case, so that  selecting upon either or both of these is an efficient way of enhancing the sensitivity to the $\theta_{12}$ and $\theta_2$ phases.

We have then classified the triple products that can be formed from the initial- and final-state momenta and the missing transverse energy vector ({\bf MET}). Upon studying the
two distributions of this kind which are not redundant, i.e., $(\mathbf{MET}\cdot (\mathbf{P}_{fe^-}\times \mathbf{P}_{fe^+}))/{s}^{3/2}$  and $(\mathbf{MET}\cdot (\mathbf{P}_{ie^-}\times \textbf{P}_{fe^-}))/{s}^{3/2}$, we were able to separate the CPV case from the CPC one, in the presence of finite resolution effects.
We stress, however, that what these distributions establish is the possibility of telling the two scenarios apart and not by itself the presence of CP-violation. 
In fact, 
the widths of the two distributions differ because the aforementioned process receives CPV contributions from $S_3$ and $S_4$ propagators (both with CP-mixed status)   whereas only the $A_1$ one (with CP-odd status) contributes in the CPC process,
all of which interfere with contributions involving the CP-even propagator of the SM Higgs state $h$. 
Establishing CP violation requires instead to extract a non-zero asymmetry between the numbers of events with positive and with negative triple product, an observable which vanishes for a CP-conserving potential (as long as the absorptive parts of the amplitude can be neglected): separating a genuine phase from such absorptive effects thus calls for further studies, in addition to the comparison with the CP-conjugate configuration.

We leave studies of the SM backgrounds for future work, as their  realistic assessment would require the inclusion of, in particular, $e^+e^-\to W^+W^-\to l^+l^-\nu\bar\nu$ and $e^+e^-\to ZZ$ events, together with a detector simulation and the construction of an integrated asymmetry with its statistical significance as a function of the integrated luminosity, all of which are beyond the scope of the present paper. Finally, we have not done a scan over the two aforementioned phases either. Rather than the comparison of isolated BPs, as done here, such a scan would  make it possible to map the region of CPV I(2+1)HDM parameter space that can actually be reached. We leave also this development for the future.

\section*{Acknowledgments}
SM is supported in part through the NExT Institute and the STFC Consolidated Grant ST/C004240/1. JH-S and CG-H acknowledge the support by SNII-SECIHTI, VIEP- BUAP and PRODEP-SEP, SECIHTI -CBF-2025-G-1187. as well as, JHS  acknowledges SECIHTI Sabbatical Support 2025 under the grant ``Higgs and Dark Matter Physics''. 

\bibliographystyle{apsrev4-2.bst}
\bibliography{bib-CPV}

\end{document}